\documentclass[preprint,double-spaced,showpacs,amsmath,amssymb]{revtex4}
\usepackage{amssymb,graphicx}
\usepackage{dcolumn}

\def\P3{{\cal P}_t}
\def\J3{{\cal J}}
\def\T3{{\cal T}}

\def\v#1{{\bf#1}}

\def\beq{\begin{equation}}
\def\eeq{\end{equation}}
\def\bar{\begin{array}[b]}
\def\barc{\begin{array}}
\def\bart{\begin{array}[t]}
\def\ear{\end{array}}

 1  1
 1 
scaled\magstephalf 
 
scaled\magstephalf

\begin{document}
\thispagestyle{empty}
\vspace*{0.5 cm}
\begin{center}
{\bf Plasmon excitations in homogeneous neutron star matter.}
\\
\vspace*{1cm} {\bf M. Baldo and C. Ducoin}\\
\vspace*{.3cm}
{\it Dipartimento di Fisica, Universit\`a di Catania}\\
and \\
{\it INFN, Sezione di Catania}\\
{\it Via S. Sofia 64, I-95123, Catania, Italy} \\
\vspace*{.6cm}
\vspace*{1 cm}
\end{center}
{\bf ABSTRACT}\\
We study the possible collective plasma modes which can affect neutron-star thermodynamics and different
elementary processes in the baryonic density range between nuclear saturation ($\rho_0$) and $3\rho_0$. In this
region, the expected constituents of neutron-star matter are mainly neutrons, protons, electrons and muons
($npe\mu$ matter), under the constraint of beta equilibrium. The elementary plasma excitations of the $pe\mu$
three-fluid medium are studied in the RPA framework. We emphasize the relevance of the Coulomb interaction among
the three species, in particular the interplay of the electron and muon screening in suppressing the possible
proton plasma mode, which is converted into a sound-like mode. The Coulomb interaction alone is able to produce
a variety of excitation branches and the full spectral function shows a rich structure at different energy. The
genuine plasmon mode is pushed at high energy and it contains mainly an electron component with a substantial
muon component, which increases with density. The plasmon is undamped for not too large momentum and is expected
to be hardly affected by the nuclear interaction. All the other branches, which fall below the plasmon, are
damped or over-damped.

\section{Introduction}
The evolution of neutron stars is in general determined by the microscopic processes that can occur in their
interior and in their crust. In particular, the cooling process by neutrino emission is strongly affected by the
detailed structure of the matter and the excitations it can sustain. Mean free path and emissivity of neutrinos
are directly related to the correlations present in the neutron star matter, which in the standard models is
composed of neutrons, protons, electron and muons~\cite{shap}. The effect of collective excitations on neutrino
emissivity have been extensively studied in the literature, often with controversial
results~\cite{Yako,Reddy,Kundu,Leinson1,Armen,Leinson2,Vosk}. The proton plasmon excitations seem to play a
relevant role in quenching the contribution of the pair-breaking mode to neutrino emissivity~\cite{Leinson1} by
charged current, since it obscures the possible effect of the Goldstone mode, which should be present in a
superfluid due to gauge invariance. However at low momentum the electron and muon screening is quite effective
and we will show that indeed the proton plasmon does not exist, and therefore its role in neutrino emissivity is
quite delicate and must be examined with care. For the same reasons, at low momenta the proton oscillations are
accompanied necessarily by electron and muon oscillations, and this can be of great relevance on the total
neutrino emission~\cite{Leinson2}. The real plasmon mode in neutron star matter is at higher energy and we will
show that it is mainly an electron excitation, with very small contribution of the proton motion but a
non-negligible muon component.
\par
All that can be best studied considering the proton, electron and muon spectral functions, that can be used also
to determine the damping of each mode. We will present extensive results on the spectral functions of the
electron-muon-proton plasma under neutron star physical conditions. Full spectrum and width of the excitations
will be presented in detail.
\par
Since neutron and proton are coupled by nuclear forces, it is not clear to what extent the presence of the
plasmon mode can affect the whole set of collective excitations or the excitation spectrum at a qualitative
level, even for non-superfluid matter. Some considerations and preliminary results will be presented on this
point, which will stress that the genuine plasma excitations are in general approximately decoupled from the
rest of the spectrum, but in particular cases they can be strongly damped.

\section{The electron-muon-proton plasma}
We consider protons, muons and electrons, in neutron star matter conditions, interacting by Coulomb coupling,
disregarding the nuclear interaction of protons with neutrons. This idealized case, which of course does not
correspond to actual physical conditions, will permit to clarify two points, also relevant for the general
physical situation. On one hand it will make more transparent the role of electron and muon screening on the
proton dynamics, on the other hand it will give clear indications on the strengths of the electrons, muons and
protons motion at the different excitation energies. The role of nuclear forces will be briefly discussed at the
end of the Section.

\begin{figure}[h]
\begin{center}
\includegraphics[width=1.0\textwidth]{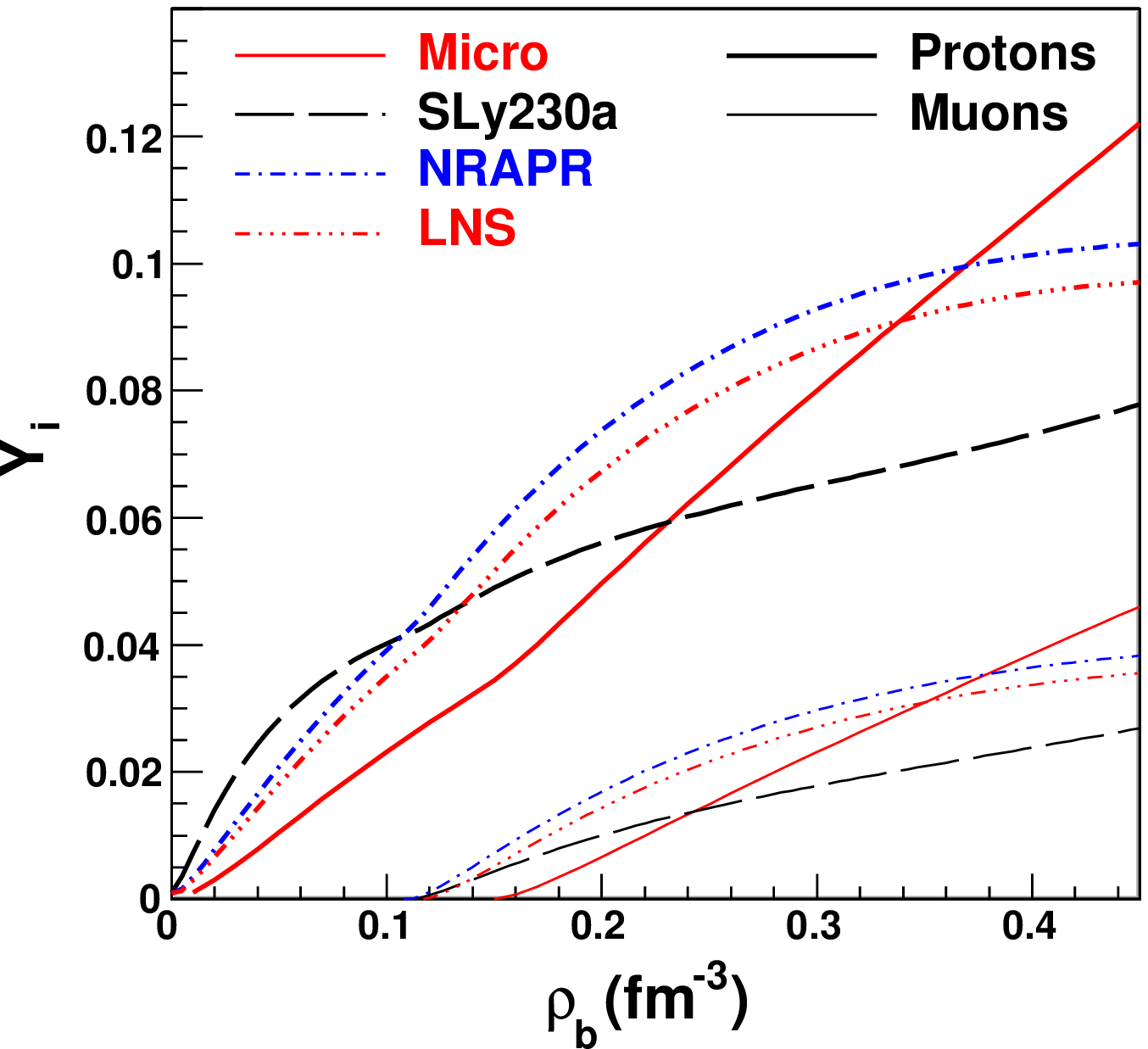}
\caption{\ }
 \label{fig:figure1}
\end{center}
\end{figure}

As already mentioned in the Introduction, we are considering normal nuclear matter, leaving the case of
superfluid matter to a future work. The range of total baryon density $\rho_b$ will be fixed between the saturation
density $\rho_0 = 0.16$~fm$^{-3}$ and $3\rho_0$, where indeed neutron star matter is expected to be homogeneous
and not affected by possible "exotic" components like hyperons.
Of course the proton fraction $Y_p=\rho_p/\rho_b$ has to be fixed in order to choose the proton density,
which is actually the only relevant parameter for the system under study.
In order to specify $Y_p$ according to neutron-star structure,
it is determined by beta equilibrium, which depends on the nuclear matter EOS.
This will be taken from the microscopic calculations of Ref.~\cite{bbb,hans}, based on the
many-body theory within the Bethe-Brueckner-Goldstone (BBG) scheme. For simplicity we will work at zero
temperature, but the results can be easily extended to finite temperature.
Electroneutrality imposes $\rho_p=\rho_e+\rho_{\mu}$,
and the proportion between electrons and muons is fixed by chemical equilibrium.
The resulting proton and muon fractions as a function of baryon density are reported in Fig.~\ref{fig:figure1}.
On this figure, the microscopic calculations are compared with results from three modern Skyrme forces:
SLy230a~\cite{SLya}, NRAPR~\cite{NRAPR} and LNS~\cite{LNS},
in order to give the order of magnitude of the dependence on the EOS.
The particle fractions used for the three baryonic densities under study are listed in Tab.~\ref{tab:Yi}.





\begin{figure}[h]
\begin{center}
\includegraphics[width=1.0\textwidth]{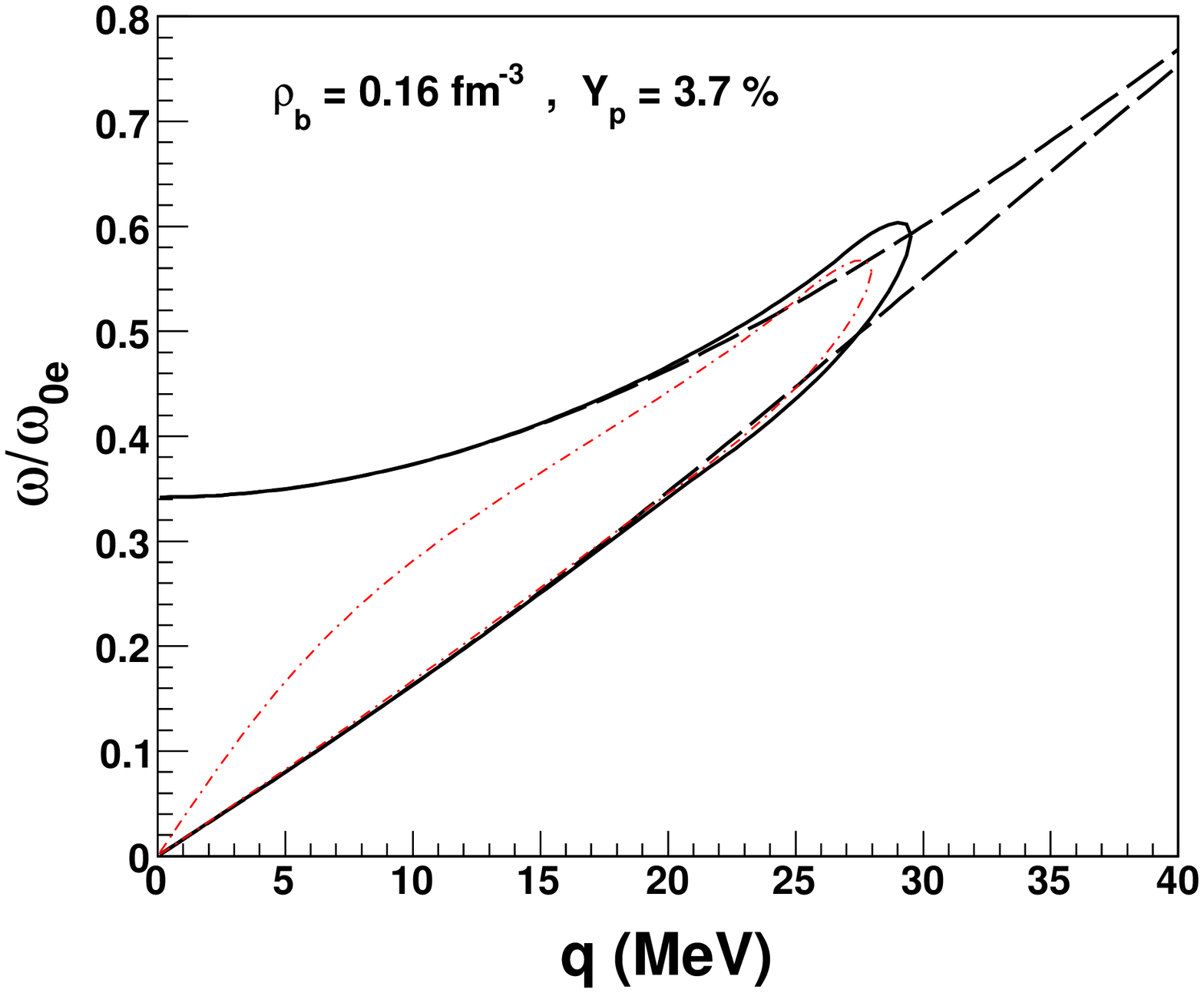}
\caption{\ }
 \label{fig:figure2}
\end{center}
\end{figure}

For a one component plasma, i.e. a charged Fermi gas on a rigid uniform background (jelly model), the Random
Phase Approximation (RPA) has been studied in detail, both in the non-relativistic~\cite{Fetter-Walecka} and
relativistic~\cite{Jancovici} cases. The excitation spectrum, i.e. the line in the energy-momentum plain where
the real part of the polarization function has poles, is known to have a "thumb like" shape. This is illustrated
in Fig.~\ref{fig:figure2} (full line), where the case of a proton gas at a density $\rho = \rho_0$ is
considered.



For future convenience, we report here the RPA equations for the polarization propagator $\Pi(k,\omega)$, which
is a function of the momentum $q$ and energy $\omega$:
\beq
\Pi(q,\omega) \, =\, \Pi_0(q,\omega)\, +\, v_c(q) \Pi_0(q,\omega) \Pi(q,\omega)\;.
\eeq
Here $v_c(q) = 4\pi e^2/q^2$ is the proton-proton Coulomb interaction and $\Pi_0(q,\omega)$ is the
free polarization propagator:
\beq
\Pi_0(k,\omega) \,=\, \int {d^3k \over {(2\pi)^3}}
\int {d\omega'\over {2\pi i}} G_0(\vert\v{k}+\v{q}\vert,\omega+\omega'))G_0(\v{k},\omega')
\eeq
and $G_0(k,\omega)$ is the free single particle Green' s function
\beq
G_0(k,\omega) = {\theta(k-k_F)\over {E-E_k + i\eta}}\, +\,{\theta(k_F-k)\over {E-E_k - i\eta}} \;,
\eeq
where $E_k$ is the single particle energy, $k_F$ the Fermi momentum and $\theta(x)$ the Heaviside step
function, which equals $1$ for $x > 0$ and is zero otherwise. The excitation spectrum is thus determined by the
equation ${\rm Re}(1 - v_c(k)\Pi_0(k,\omega)) = 0$. The explicit expression for the free polarization in the
non-relativistic limit can be found in textbooks and is usually referred to as the Lindhard function. The upper
branch appearing in Fig.~\ref{fig:figure2} is called the plasmon excitation, which is characterized by a finite
energy at vanishing momentum and has a simple classical interpretation~\cite{Rax}. The lower branch is actually
strongly damped and does not correspond to a real excitation which is able to propagate. This point will be
further elaborated in the sequel. From the Figure it is clear that in any case there is a maximum value of the
momentum above which the spectrum stops and no excitation is possible. This is mainly a quantal effect, which is
not present in the semi-classical approximation of RPA, which is nothing else that the Vlasov approximation.
This is illustrated in the same Fig.~\ref{fig:figure2}, where the RPA and the Vlasov approximation (dashed
lines) spectra are compared. Apart from the momentum cutoff, one can see that the spectra are quite similar.
The results for the relativistic case are analogous~\cite{Jancovici}, except that the "thumb" becomes thinner
and thinner as the Fermi momentum becomes more and more relativistic~\cite{McOrist}. In the ultra-relativistic
case the two branches become extremely close near the cutoff, and the difference  between RPA and Vlasov
spectrum appears smaller and smaller, apart again for the presence of the momentum cutoff.
In the following, all the calculations will be performed in the Vlasov approximation
for the three fluids, protons, electrons and muons.

\begin{figure}[h]
\begin{center}
\includegraphics[width=1.0\textwidth]{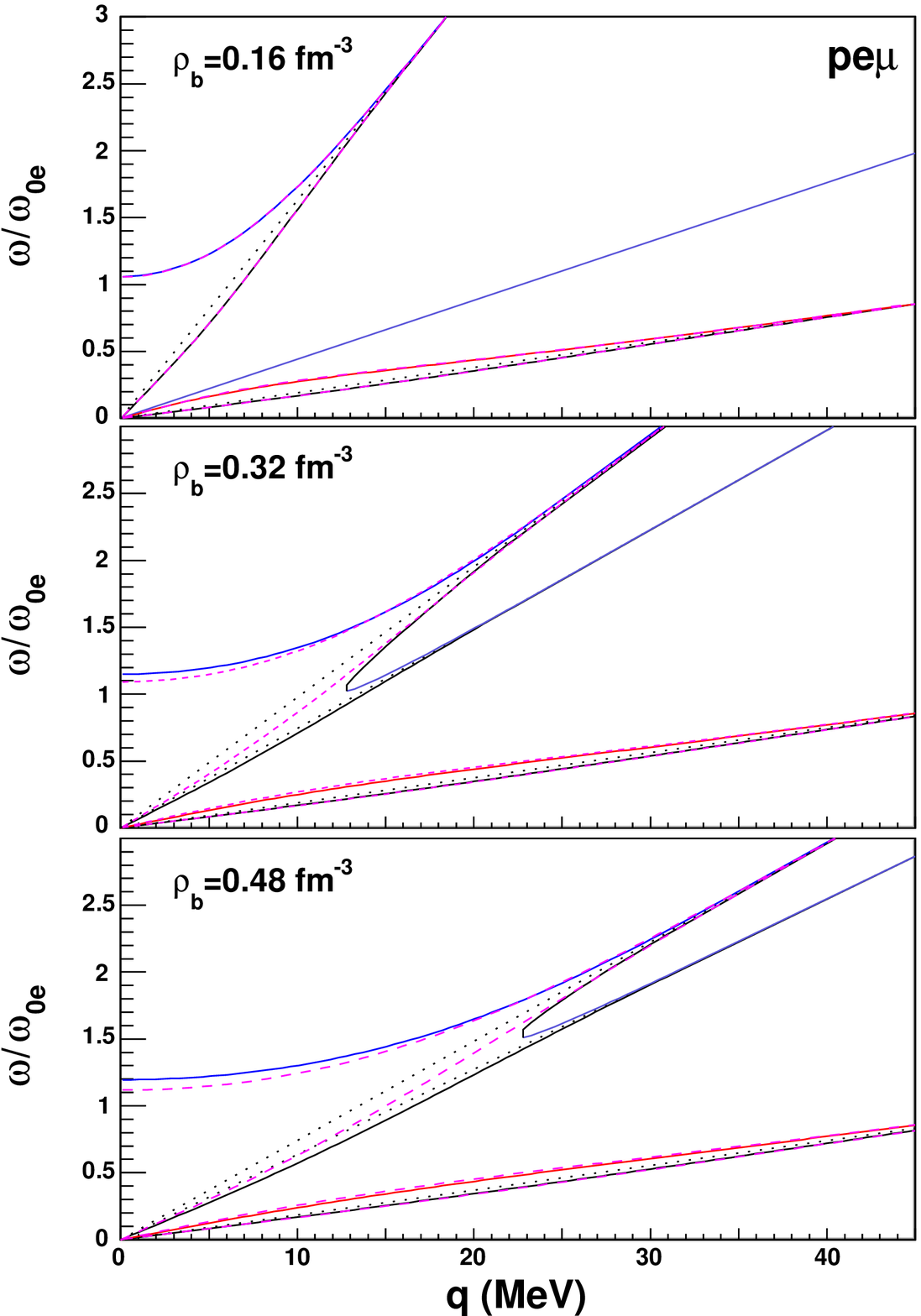}
\caption{\ }
\label{fig:figure3}
\end{center}
\end{figure}

Having summarized the well established results for the excitation spectrum with the pure Coulomb interaction
within the jelly model, let us consider the system of protons, electrons and muons, treating  the three
components on equal footing. The RPA equations for this case are simply a three by three system, where the
coupling is again provided by the Coulomb interaction between protons, muons and electrons:
\beq
\label{eq:ep}
\Pi^{ik}(q,\omega) \, =\, \Pi_0^{ii}(q,\omega)\left(\delta_{ik}\, +\,\, \Sigma_{j=e,\mu,p}\,\, v_c^{ij}(q)
\Pi^{jk}(q,\omega) \, \right) \;.
\eeq
The polarization propagator $\Pi^{ik}$ is now a three by three matrix. In Eq.~(\ref{eq:ep}) the index
$i, j, k$ stands for electrons ($e$), muons ($\mu$) and  protons ($p$).
These equations describe the motion of a three-component fluid.
It has to be stressed that no exchange is included, i.e. only density-density
correlations are considered. Since both electrons and muons are expected to be faster than protons, they are
able to follow their motion and are more effective in screening the proton-proton interactions. The total
spectrum is now formed by several branches, but only one is expected to have the characteristic property of a
finite excitation energy at vanishing momentum, and this will be identified with the plasmon mode of the
proton-muon-electron coupled system. The original proton plasmon mode becomes now a sound-like mode, since, due
to the electron and muon screening, the effective proton-proton interaction becomes of finite range (screened
Coulomb interaction). This is a well known phenomenon also in classical plasma~\cite{Rax}. The screening can be
easily seen in the present case by solving Eqs.~(\ref{eq:ep}) for the proton-proton polarization propagator.
After a short algebra one finds:
\beq
 \left( 1\, -\, \Pi_0^{pp}{v_{\rm{c}}\over {1 - \left(\Pi_0^{ee}+\Pi_0^{\mu\mu}\right)v_{\rm{c}} }}\right)
 \Pi^{pp}\, =\, \Pi_0^{pp}\;.
\eeq
Since the unperturbed electron and muon plasmon modes are substantially higher than the corresponding proton
one, in the small $q$ limit one can expand the free electron polarization propagator at zero frequency and one
gets
\beq
 \label{eq:stat}
 \left( 1\, -\, \Pi_0^{pp}{4\pi e^2\over {q^2 + q_c^2}}\right) \Pi^{pp}\, =\, \Pi_0^{pp}\;,
\eeq
which is an RPA equation with a screened Coulomb interaction and therefore of finite range.
For $q< q_c$ the collective mode can be only a sound mode with a typical linear dependence on momentum.
The inverse $\lambda = 1/q_c$ is the screening length.
One finds that $ q_c^2 = 3[(\omega_p^e/v_F^e)^2 + (\omega_p^\mu/v_F^\mu)^2]$,
where $\omega_p^e$ and $\omega_p^\mu$ are the electron and muon gas plasmon frequencies,
respectively, and $v_F^e$, $v_F^\mu$ the corresponding Fermi velocities. In neutron star matter conditions $q_c
<< k_F^p$ and the screening length is much larger than the average distance between protons. Therefore, at
increasing momenta the proton excitation should merge in the proton plasmon mode with no screening. Besides
that, the static limit we have performed in Eq.~(\ref{eq:stat}) is only approximately valid, especially at the
higher density, where the proton plasma frequency is a substantial fraction of the electron plasma frequency,
and it is more appropriate to solve explicitly Eq.~(\ref{eq:ep}). The actual calculations confirm these
expectations, as illustrated in Fig.~\ref{fig:figure3} for three proton densities corresponding to total baryon
densities $\rho_0$, 2$\rho_0$ and 3$\rho_0$.
The reported branches are obtained by searching for the energies for which the determinant $\Delta$ of the real
part of the three by three matrix in Eq.~(\ref{eq:ep}) vanishes at a given momentum.
In the present particular case it has a simple expression:
\beq
 \Delta\, =\, 1\, -\, v_c (\Pi_0^{ee}\, +\,\Pi_0^{\mu\mu}\, +\, \Pi_0^{pp})
 \label{eq:det}
\eeq
One observes that all branches go to zero at vanishing momentum, except the upper one. The latter is
the plasmon mode of the three components system, while the "thumb like" shape of the original proton branch in
the uncoupled system can be identified with the lowest "loop" (color on line). Additional branches appear in
between these two. For $\rho = \rho_0$ one additional branch approaches closely the plasmon one at increasing
momentum. It can be considered as the original "sound like" electron mode, according to the above discussion for
the proton gas in the jelly model. It turns out, as we will show below, that it is still mainly an electron
excitation even in the coupled case. The last additional excitation, which looks also a "sound like" mode is
reminiscent of a pure muon excitation, which however is now mixed with a small electron component. This coupling
become stronger at increasing density and the two branches merge in a single one with a typical "bending back",
see panels for $\rho = 2\rho_0$ and $\rho = 3\rho_0$, which actually marks the smooth transition form almost a
pure electron excitation to a mixed electron-muon excitation.



\begin{figure}[h]
\begin{center}
\includegraphics[width=1.0\textwidth]{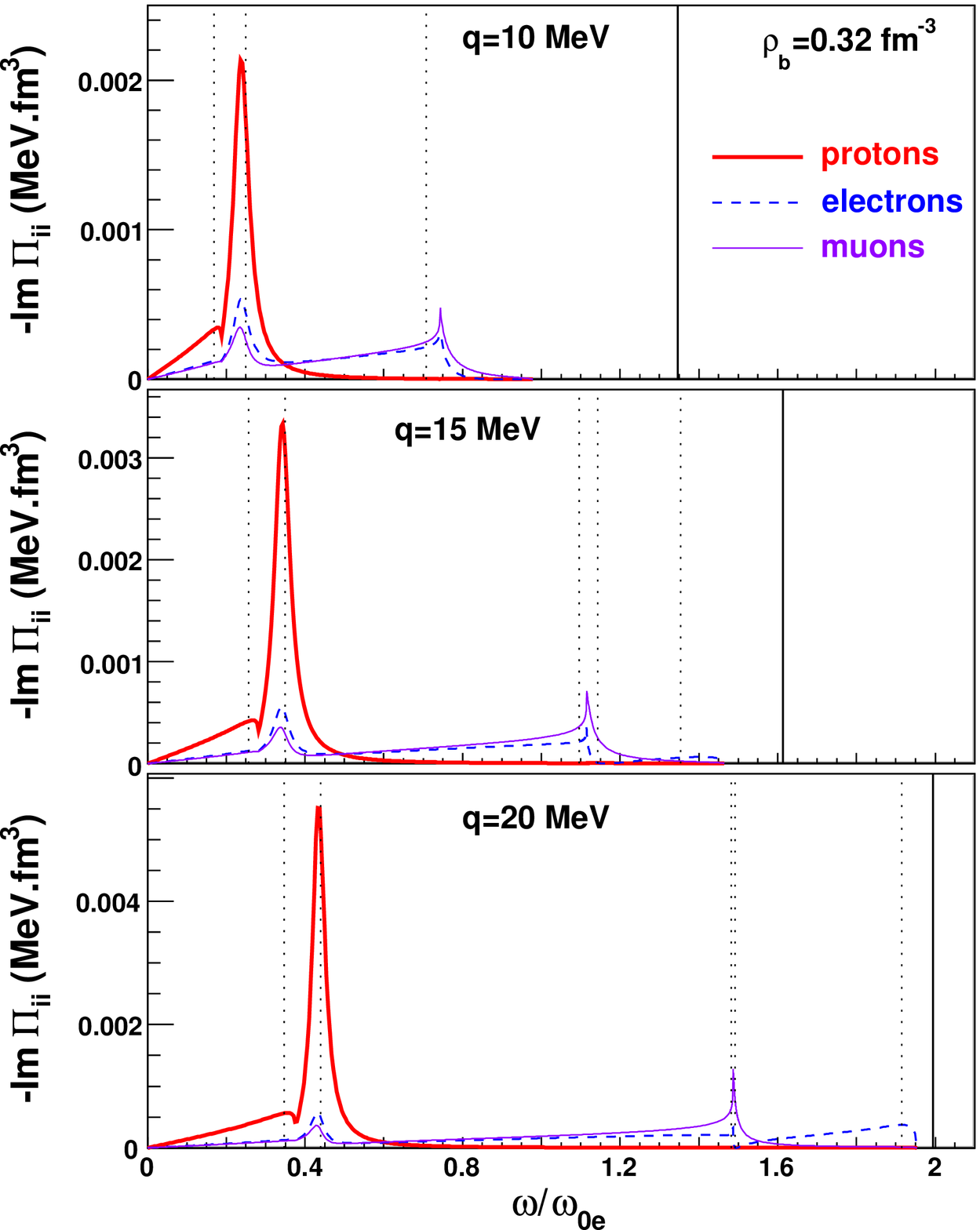}
\caption{\ }
\label{fig:figure4}
\end{center}
\end{figure}

To support and illustrate this type of analysis we have reported in the same figure the branches (dashed lines)
one obtains when one considers only protons and electrons.
In this case, we keep the original proton density
and substitue the electron-muon liquid by a pure electron liquid with $\rho_e=\rho_p$.
From the figure one can trace the evolution of the branches when introducing the muon component.
It has to be noticed that all branches, except the plasmon one, are actually damped.
The polarization propagators $\Pi_0$ which appear in
Eqs.~(\ref{eq:ep}) are indeed in general complex. The best way to estimate the damping rate of each branch is to
calculate the spectral function $-{\rm Im}(\Pi^{ii}(q,\omega))$, which, in the present case, has a proton
($i=p$), an electron ($i=e$) and a muon (i=$\mu$) component. This quantity is also the basic one for calculating
the rate of many physical processes. At a given momentum $q$, as a function of $\omega$ this quantity is
expected to display peaks corresponding to the different branches, whose width gives a measure of their damping
rate, i.e. each excitation mode is actually a resonance if the width is not too large. The plasmon, up to a
certain momentum, has no width, and the strength function presents a delta function singularity at the plasmon
frequency. The electron, muon and and proton strength functions are reported in Fig.~\ref{fig:figure4} at twice
the saturation density and at selected values of the momentum $q$. For guidance, a dashed vertical line is
reported at the position of each branch. To put in evidence the plasmon branch we have reported just a sharp bar
at the position of its energy. One can see that that peaks are present at the energy corresponding to the
branches of Fig.~\ref{fig:figure3}, even if some slight shifts occur.
Such shifts are expected, since the presence of an imaginary part has also the effect to produce some displacements.



The relatively narrow peak at lower energy corresponds to the original proton plasmon in the uncoupled system.
As discussed above, this branch in the coupled system is a sound-like mode due to electron and muon screening.
For this branch the proton component is larger, but electrons and muons follow the proton oscillation, which
allows a quite effective screening. However, as the momentum increases the screening becomes less effective, in
agreement with the discussion above. Besides that, the coupling with electrons and muons produces a substantial
damping of this branch. At higher energy a small broad peak is obtained, which involves a substantial fraction
of both muons and electrons, but not protons. This branch is a coupled oscillations of electrons and muons,
which behave approximately as a single fluid.
It is characteristic of the coupling between electrons and muons,
and would not be present in a system with only protons and electrons.
In any case, as one can see,
the strength associated with this structure is quite low.
It has to be noticed that it is present also at $q =10$ MeV,
where in principle no branch is present at this energy. The fact that this structure does not
correspond to the vanishing of the real part of the RPA determinant is an indication that the corresponding
oscillation is over-damped, and indeed it is quite broad at this momentum.

The very broad structure which appears just below the sharp plasmon peak is only due to the electrons, and it
corresponds to the over-damped sound-like branch which is present in a single fluid, below the plasmon, as
discussed for the proton case in relation with Fig.~\ref{fig:figure2}. Its strength is increasing at increasing
momentum.

\begin{figure}[t]
\begin{center}
\includegraphics[width =0.8\linewidth]{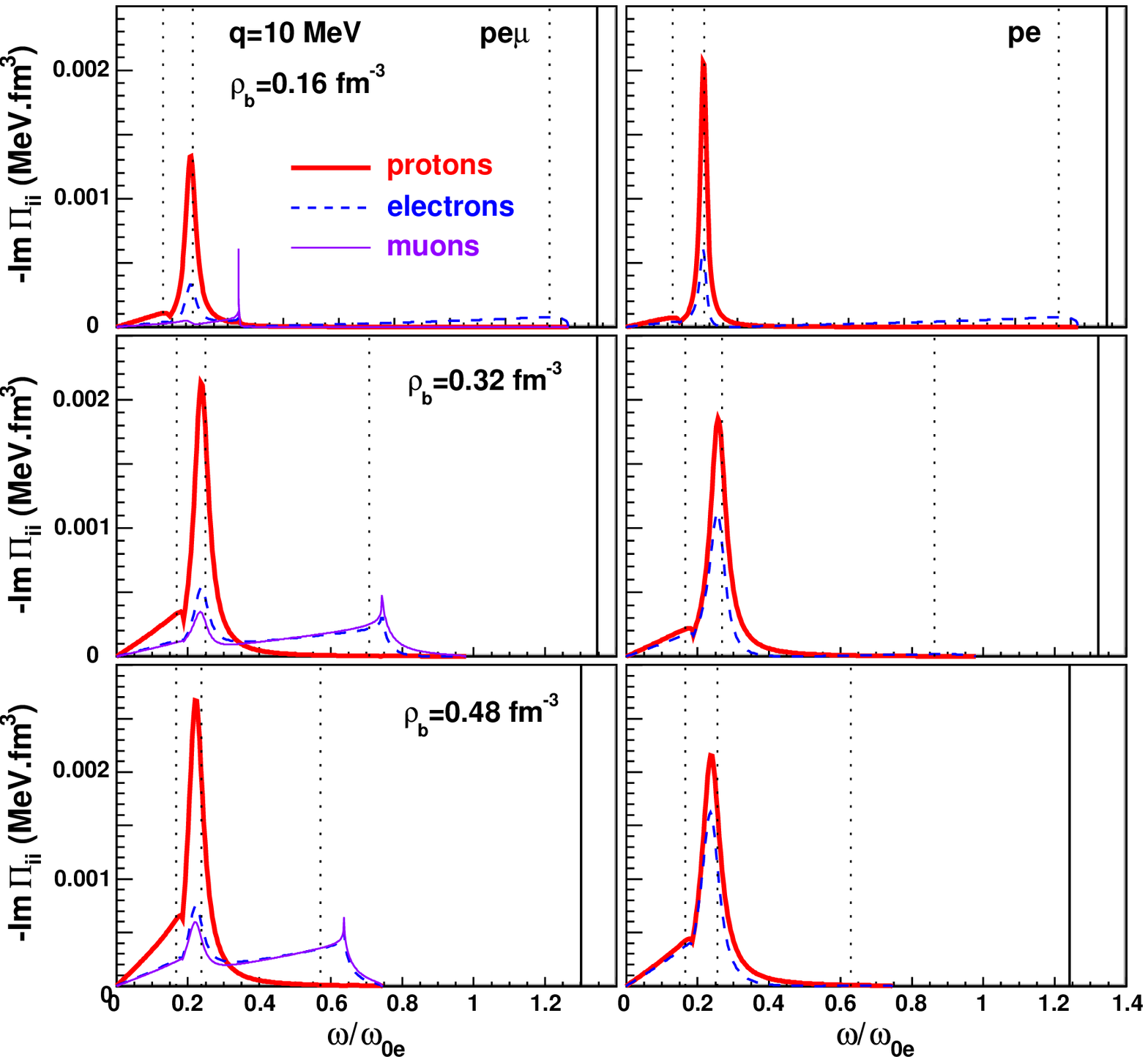}
\end{center}
\caption{\ } \label{fig:figure5}
\end{figure}

To illustrate and support these considerations and interpretations of the results,
we compare in Fig.~\ref{fig:figure5} at different densities and for $q = 10$ MeV the strength functions of the
proton-muon-electron system with the ones of the two-component proton-electron system, where the electron
density equals the electron-muon density of the three-component system. One can see that the electrons and muons
move together in screening the proton plasmon, with a total strength not so far from the electron strength in
the two-component system. As anticipated above, the broad electron-muon structure is not present in the
two-component system, while the broad electron structure below the plasmon peak is mainly unaffected.
However, at the lowest density one notices a muon peak just above the proton sound-like mode
and a very weak involvement of the muons in the screening of the proton motion.
The muon peak corresponds to the branch at "intermediate energy" in Fig.~\ref{fig:figure3}.
This means that at this density the muons are not able to follow the proton motion
and produce an independent mode, which however has not a plasmon-like character since its energy
is mainly linear in momentum.



Let us analyze in more detail the plasmon mode, which is the highest in energy and appears in the spectral
function as a delta-function singularity. In fact at the plasmon energy all the free propagators $\Pi_0^{ii}$
are real. Solving the RPA equations, one gets:
 \beq
 \label{eq:pi}
 \Pi^{ii}\, =\, { {\Pi_0^{ii}[ 1 - (\Pi_0^{jj} + \Pi_0^{kk})v_c]}\over {\Delta + i\eta} }\;,
\eeq
where $i \neq j \neq k$ and $\eta$ is the usual infinitesimal quantity demanded by the general
structure of the response function. The plasmon contribution to the spectral function is then:
\beq
 \label{eq:plas}
-{\rm Im}(\Pi^{ii}) \, =\, v_{\rm c}(\Pi_0^{ii})^2 \delta(\omega - \omega_p)/|\Delta(\omega_p)'|
\eeq
where $\omega_p$ is the plasmon energy, solution of the equation $\Delta(\omega)
= 0$. The ratios of the different strengths are then given by
\beq
\label{eq:str}
 { {\rm Im}(\Pi^{ii}) \over {\rm Im}(\Pi^{jj}) }  \, =\, { { (\Pi_0^{ii})^2 }\over { (\Pi_0^{jj})^2 }  }\;,
\eeq
where the expression of Eq.~(\ref{eq:det}) for the determinant has been used. The plasmon pole corresponds also
to the zero eigenvalue of the RPA matrix which in Eq.~(\ref{eq:ep}) defines the response function $\Pi(\omega)$,
i.e.
\begin{equation}
\label{eq:eig}
      \Sigma_{j=e,\mu,p}\,\, \left(\delta_{ij}\, -\,\, \Pi_0^{ii}(\omega_p) v_c^{ij}(q) \right)
      \delta\rho_j \, =\, 0\;,
\end{equation}
where the eigenvector $\delta\rho = (\delta\rho_e , \delta\rho_\mu , \delta\rho_p)$ describes the
electron, muon and proton density fluctuations of the modes.
One easily finds
\begin{equation}
\label{eq:ro}
  {\delta\rho_i\over \delta\rho_j}\, =\, z_i z_j {\Pi_0^{ii}\over \Pi_0^{jj}}\;,
\end{equation}
where $z_i = \pm 1\,\,$ is the charge sign of the $i$-component.
This result is in agreement with Eq.~(\ref{eq:str}), since the spectral function gives the {\it square} of the density fluctuation.

\begin{figure}[t]
\begin{center}
\includegraphics[width =0.8\linewidth]{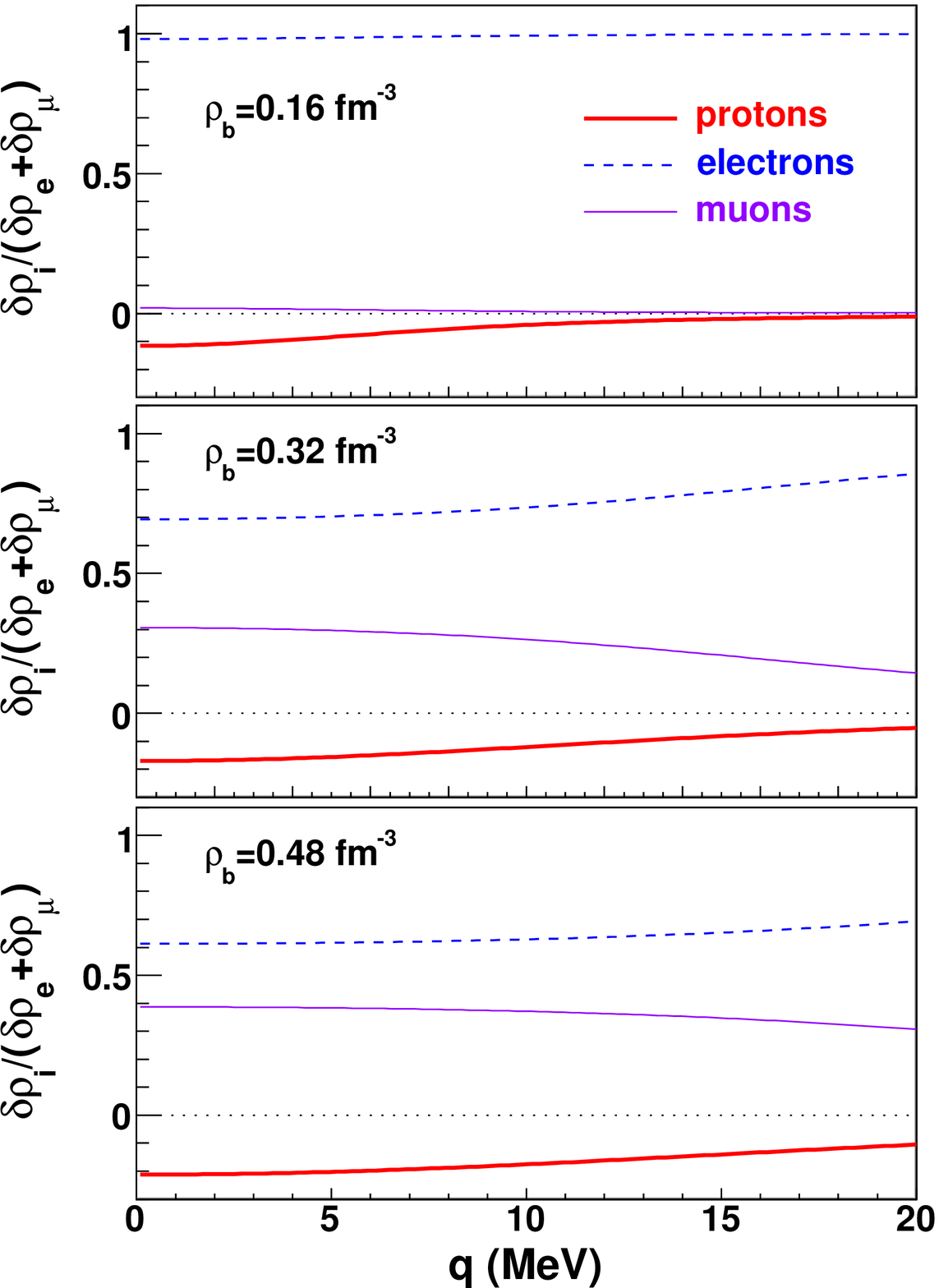}
\end{center}
\caption{\ } \label{fig:figure6}
\end{figure}

In Fig.~\ref{fig:figure6} are reported some ratios of the density fluctuations,
according to Eq.~(\ref{eq:ro}), which describes the plasmon mode at different momenta.
The main characteristic to be noticed is the dominance of the
electron component, with the muons becoming of some relevance only at the higher densities. The proton component
is marginal, since protons are unable to follow the rapid motion of the negatively charged components. The
momentum dependence is quite weak.



In the analysis we have carried out, the nuclear interaction and the presence of neutrons have been completely
neglected. However, the plasmon mode is at relatively large frequency, while the nuclear sound (or zero-sound)
mode starts linearly with momentum. The coupling between the plasmon and the other modes is expected to be quite
weak, and the properties of the plasmon excitation are likely to be essentially unaffected.
Furthermore, the plasmon mode is known not to be influenced by the possible presence of a superfluid phase~\cite{Schri}.

The nuclear interaction can affect the lower energy modes, whose strength will be then redistributed.
In particular the sound-like mode involving the protons will be coupled with neutron modes, since it falls in
the low frequency part of the spectrum, due to the screening effect discussed above. This show the relevance of
the Coulomb interaction for the overall structure of the spectrum. A complete analysis of the excitation
strength distribution, including both Coulomb and nuclear interaction, will be given elsewhere.

\section{Conclusion and prospects.}
We have analyzed, within the RPA scheme, the plasmon excitations which can occur in the homogeneous matter
present in the outer core of Neutron Stars. The proton, electron and muon components have been considered. Since
both electron and muon Fermi velocity are larger than the proton one, the Coulomb proton-proton interaction is
effectively screened and no genuine proton plasmon can exist, but instead a sound-like mode is present. The
analysis of the spectral functions shows that the high frequency plasmon mode is mainly an electron excitation,
with possibly a substantial muon component at higher density. Except for the genuine plasmon mode, all other
excitation modes are damped, and in some case over-damped. The inclusion of the nuclear interaction and of the
neutron component will be reported elsewhere.

\newpage

\begin{table}[t]
\begin{center}
\begin{tabular}{|c|c|c|c|c|}
\hline
$\rho_{\rm b}$ $[{\rm fm}^{-3}]$&$Y_p$ $[\%]$ &$Y_e$ $[\%]$ &$Y_{\mu}$ $[\%]$ & $\omega_{0e}$ $[MeV]$\\
\hline
$0.16$ & $3.70$ & $3.63 $ & $0.07 $ & $6.15 $ \\
\hline
$0.32$ & $8.57$ & $5.95 $ & $2.62 $ & $10.25 $ \\
\hline
$0.48$ & $13.03$ & $7.98 $ & $5.05 $ & $13.48 $ \\
\hline
\end{tabular}
\end{center}
\caption{ Particle fractions for the baryonic densities under study. The electron plasmon frequency for
$\rho_e=\rho_p$, used as a unit for the plots, is also given. } \label{tab:Yi}
\end{table}%

\noindent Figure captions.

\par

\noindent Fig. 1 - The proton fraction (thick lines) and muon fraction (thin lines) as a function of total
baryon density in neutron star matter. Microscopic calculations (taken as a reference) are compared with results
from modern Skyrme forces.

\vskip 0.25 cm
\par

\noindent Fig. 2 - The proton excitation spectra in the jelly model. The energy $\omega_{0e}$ is the electron
plasmon energy at zero momentum, for $\rho_e=\rho_p$. The full line is obtained in the RPA scheme, the
long-dashed line corresponds to the Vlasov approximation. The dashed-dotted line shows the effect of electron
screening on the (RPA) proton spectrum.

\vskip 0.25 cm
\par

\noindent Fig. 3 - Dispersion relation in the $pe\mu$ system (full lines), compared with the $pe$ system (dashed
lines), for three baryonic densities. The corresponding particle fractions are given in Tab.~\ref{tab:Yi}. The
dotted lines correspond to $\omega_i=v_{{\rm F}i}q$. The energy $\omega_{0e}$ is the electron plasmon energy for
$\rho_e=\rho_p$.

\vskip 0.25 cm
\par

\noindent Fig. 4 - Diagonal elements of the spectral-function matrix ($pe\mu$ system), for three momenta and
$\rho_{\rm b}=0.32$ fm$^{-3}$. The vertical lines indicate the energy at which the real part of the determinant
of Eq. \ref{eq:det} vanishes. For the plasmon the dashed and full lines coincide. The energy $\omega_{0e}$ is
the electron plasmon energy for $\rho_e=\rho_p$.

\vskip 0.25 cm
\par

\noindent Fig. 5 - Diagonal elements of the spectral-function matrix ($pe\mu$ system), for three different
densities and $q=10$ MeV. The energy $\omega_{0e}$ is the electron plasmon energy for $\rho_e=\rho_p$. The
vertical lines indicate the energy at which the real part of the determinant of Eq.~\ref{eq:det} vanishes. For
the plasmon the dashed and full lines coincide. Left: $pe\mu$ system; right: $pe$ system.

\vskip 0.25 cm
\par

\noindent Fig. 6 - Direction of the eigenvector associated with the plasmon mode.

\par

\par

\par

\par

\par

\end{document}